# THE ANTI-EINSTEIN EQUATIONS

**Evangelos Chaliasos, PhD**

*Research Center for Astronomy and Applied Mathematics (RCAAM)*
*Academy of Athens*

*4 Soranou Efessiou, GR-11527 Athens, GREECE*
chaliasose@gmail.com

*Abstract*

As we know, from the Einstein equations the vanishing of the four-divergence of the (symmetric) energy-momentum tensor follows. This is the case because the four-divergence of the Einstein tensor (which is also symmetric) vanishes identically. Inversely, we find that from the vanishing of the four-divergence of the energy-momentum tensor (consisting in the general case of both symmetric and antisymmetric parts) not only the Einstein equations follow (concerning the symmetric part), but also the so-named anti-Einstein equations (concerning the antisymmetric part). These equations must be considered as complementary to the Einstein equations. Also, while from the Einstein equations the energy density (or the pressure) can be found (without using an equation of state), from the vanishing of the four-divergence of the (symmetric) energy-momentum tensor the pressure (or the energy density) can also be found, without having to use an additional (but arbitrary) equation of state.

1.  **Introduction**

As we know [3], the Einstein equations are ten. In order to solve them, we have also available four degrees of gauge freedom, resulting in fixing four of the components of the metric tensor by means of four coordinate transformations available. Thus it is sufficient to find from the Einstein equations only six of the ten components of the metric tensor. Then we will have to find four additional unknowns. These are three components of the four-velocity (the fourth component being fixed by the normalization condition for the four-velocity), and the density (or the pressure) of the matter, the pressure (or the density) being fixed by introducing an extra (arbitrary) equation, the equation of state. However we can get rid of this arbitrariness, if we take as the extra equation the equation expressing the vanishing of the four-divergence of the (symmetric) energy-momentum tensor (resulting from the Einstein equations), which then can substitute the *equation of state.* Also, as we will see in what follows, from the integration of the equations expressing the vanishing of the four-divergence of the *total* energy-momentum tensor not only the Einstein equations result. In fact the *Einstein* equations are merely the *symmetric* part of the resultant equation. And the *antisymmetric* part constitutes what we may call the *anti-Einstein* equations.

## 2. The anti-Einstein equations

**I.** First of all, I have to mention that the energy-momentum tensor $T^{ik}$ in general is *not* necessarily symmetric, but we can make it symmetric by a standard procedure (see [3], § 32).

If $G^{ik}$ is the Einstein tensor, defined by

$$G^{ik} = R^{ik} - (1/2)g^{ik}R, \tag{1}$$

the Einstein equations are given by

$$G^{ik} = T^{(ik)}. \tag{2}$$

Here $T^{(ik)}$ is the *symmetric* part of the energy-momentum tensor $T^{ik}$, defined by

$$T^{(ik)} = (1/2)\left(T^{ik} + T^{ki}\right), \tag{3}$$

while $T^{[ik]}$ is its *antisymmetric* part, defined by

$$T^{[ik]} = (1/2)\left(T^{ik} - T^{ki}\right). \tag{4}$$

Because

$$G^{ik}_{;k} = 0 \tag{5}$$

identically, we have from (2)

$$T^{(ik)}_{;k} = 0. \tag{6}$$

But (6) contains the symmetric part of the equations of motion, and also is contained in the field equations (2) [3] (p. 277). It has to be noted however that, in view of (6), equation (5) has to be integrated rather than the Einstein equations (2), because it contains accelerations, while (2) contains only velocities. Note also that, more generally than (6),

$$T^{ik}_{;k} = 0, \tag{7}$$

which expresses the equations of motion of matter plus fields other than gravitational, as being essentially the Lagrange equations for them (see [3], § 32) in *curved* spacetime. From (7) and (6) now we also must have

$$T^{[ik]}{}_{;k} = 0. \tag{8}$$

Integrating Eq. (7) we find that $G^{ik} - T^{ik}$ must be equal to a tensor field whose divergence is zero. Such a tensor field must be the **curl** of a vector field $X_m$ (say). The **curl** of a vector field $X_m$ is defined as

$$E^{iklm} X_{m;l}, \tag{9}$$

where

$$E^{iklm} \equiv \left(1/\sqrt{-g}\right) e^{iklm} \tag{10}$$

is the completely antisymmetric tensor, with $e^{iklm}$ the completely antisymmetric symbol.[*] If thus we separate *(uniquely)* $T^{ik}$ to symmetric and antisymmetric parts according to

$$T^{ik} = T^{(ik)} + T^{[ik]}, \tag{11}$$

we get from (7), because also of (5),

$$G^{ik} - T^{(ik)} = T^{[ik]} + E^{iklm} X_{m;l}, \tag{12}$$

since the divergence of a curl (of $E^{iklm} X_{m;l}$) gives 0. In fact,

$$\begin{aligned}
\left(E^{iklm} X_{m;l}\right)_{;k} &= E^{iklm} X_{m;l;k} \\
&= E^{ilkm} X_{m;k;l} \\
&= -E^{iklm} X_{m;k;l} \\
&= (1/2) E^{iklm} (X_{m;l;k} - X_{m;k;l}) \\
&= (1/2) E^{iklm} X_p R^p{}_{mlk}, \tag{13}
\end{aligned}$$

where the last equality holds $\forall$ $X_m$ by virtue of the properties of the Riemann tensor [3]. But

$$\begin{aligned}
e^{iklm} X_p R^p{}_{mlk} &= -e^{iklm} X_p R^p{}_{klm} \\
&= -e^{iklm} g_{pr} X^r R^p{}_{klm} \\
&= -e^{iklm} \delta^p{}_r X^r R_{pklm} \\
&= -X^r (e^{iklm} \delta^p{}_r R_{pklm}). \tag{14}
\end{aligned}$$

Now, the quantity inside the parenthesis is

$$e^{iklm}\delta^p_{\ r}R_{pklm} = e^{iklm}R_{rklm}$$
$$= e^{rklm}R_{rklm}, \tag{15}$$

where there is *no* summation with regard to r (being equal to i). Then the result is null by twice applying the cyclic identity. Thus, finally, from (13), (14), &(15) we get

$$(E^{iklm}X_{m;l})_{;k} = 0, \tag{16}$$

which is an identity with respect to $X_m$ (div curl $X_m = 0$ ∀ $X_m$).

---

[*] To compute $E^{iklm} X_{m;l}$, we have to observe that this is equal to $E^{ikml} X_{l;m} = -E^{iklm} X_{l;m}$, and thus it equals $(1/2) E^{iklm} (X_{m;l} - X_{l;m}) = (1/2) E^{iklm} (\partial X_m/\partial x^l - \partial X_l/\partial x^m)$.

**II.** But an important remark has to be made here concerning the equation (12). Looking at (12), we observe that the left hand side is symmetric, while the right hand side is antisymmetric. Thus, in order for (12) to hold, it is necessary for both sides to vanish. We get therefore the equations

$$G^{ik} - T^{(ik)} = 0 \quad \& \tag{17}$$
$$T^{[ik]} + E^{iklm}X_{m;l} = 0. \tag{18}$$

We recognize at once (17) as the *Einstein equation (symmetric),* while (18) is what may be called the *anti-Einstein equation,* because it is *antisymmetric.* Obviously, when $T^{[ik]}$ degenerates, I mean when $T^{[ik]} = 0$, then (18) has as a solution $X_m$ the gradient of any scalar field (curl grad φ = 0  ∀ φ )[**].

### 3. $T^{ik}_{\ ;k} = 0$ as the equation of state

As we have already said (Eq. (7)),

$$T^{ik}_{\ ;k} = 0 \tag{19}$$

can be taken as the equation of state. I will show it with a specific example, namely by examining in detail the case of a perfect fluid, whose energy-momentum tensor (symmetric) is given by

$$T^{ik} = (p+\varepsilon)u^i u^k - pg^{ik}. \tag{20}$$

First, because it will be needed in the sequel, I will find the quantity $(u^i u^k)_{;k}$. Thus, starting from the identity $u^i u_i = 1$, we will have $(u^i u_i)_{;k} = 0$. As a consequence

$$u^i_{\ ;k} g_{il} u^l + u^i u_{i;k} = 0, \tag{21}$$

or

$$u_{i;k}\delta^i_{\ l} u^l + u^i u_{i;k} = 0 \tag{22}$$

---

[**] $E^{iklm} X_{m;l} = 0 \Rightarrow^{(*)} \partial X_m/\partial x^l - \partial X_l/\partial x^m = 0 \Rightarrow X_m = \varphi_{,m}$ because $\varphi_{,m,l} = \varphi_{,l,m}$ (Schwartz's theorem).

or

$$u_{l;k}u^l + u^i u_{i;k} = 0, \quad (23)$$

or

$$u^i u_{i;k} = 0. \quad (24)$$

We have thus

$$0 = T^k_{i\ ;k} = \left[(p+\varepsilon)u_i u^k - p\delta^k_i\right]_{;k}, \quad (25)$$

from which

$$p_{;i} = \left[(p+\varepsilon)u_i u^k\right]_{;k} = (p_{;k} + \varepsilon_{;k})u_i u^k + (p+\varepsilon)(u_i u^k)_{;k}. \quad (26)$$

Multiplying both sides by $u^i$, we find

$$0 = p u_{i;k} u^i u^k + p u^k_{;k} + \varepsilon_{;k} u^k + \varepsilon u_{i;k} u^i u^k + \varepsilon u^k_{;k}, \quad (27)$$

and taking in mind (24)

$$0 = p u^k_{;k} + \varepsilon_{,k} u^k + \varepsilon u^k_{;k}, \quad (28)$$

so that finally

$$p + \varepsilon = -\varepsilon_{,k} u^k / u^k_{;k}. \quad (29)$$

This is the equation of state.

Note that this gives

$$p + \varepsilon = 0 \quad (30)$$

in the case of the *perfect* cosmological principle ($\varepsilon_{,k} = 0\ \forall\ k$) (*steady state cosmology*). But also cf. [1] & [2], where we find that (30) also holds in the case of a *rotation* of the Universe.

*Acknowledgement:* I wish to thank Professor George Contopoulos for some valuable suggestions.

## REFERENCES

[1] Chaliasos, E. (2006): «The Rotating and Accelerating Universe», arXiv.org/abs/astro-ph/0601659

[2] Chaliasos, E. (2011): «Cosmological Results from a Newtonian Rotation of the Universe», ABI Chronicles, Raleight, N.C., USA

[3] Landau, L.D. & Lifshitz, E.M. «The Classical Theory of Fields», Vol. 2 of «Course of Theoretical Physics» 1975 (4th ed.), Pergamon